\documentclass[epj]{svjour}
\usepackage{graphicx}
\newcommand{\cd}{\makebox[0.08cm]{$\cdot$}}
\newcommand{\sla}{\not\!}
\begin{document}
\title{The pion wave function in covariant light-front dynamics}
\subtitle{Application to the calculation of various physical observables}
\author{O. Leitner \inst{1} \and J.-F. Mathiot \inst{2} \and N. Tsirova \inst{2}   
}                     
\institute{Laboratoire de Physique Nucl\'eaire et de Hautes \'Energies,\ 
Groupe Th\'eorie, Universit\'e Pierre et Marie Curie et Universit\' e Diderot, CNRS/IN2P3,
4 place Jussieu, F-75252 Paris, France
\and 
Clermont Universit\'e, Laboratoire de Physique
Corpusculaire, CNRS/IN2P3, BP10448, F-63000 Clermont-Ferrand, France
}
\abstract{
The structure of the pion wave function in the relativistic constituent quark model is investigated in the explicitly covariant formulation of light-front dynamics. We calculate the two relativistic components of the pion wave function in a simple one-gluon exchange model and investigate various physical observables: decay constant, charge radius, electromagnetic and transition form factors. We discuss the influence of the full relativistic structure of the pion wave function for an overall good description of all these observables, including both low and high momentum scales.
} 
\maketitle
%
\section{Introduction}
%
The understanding of the internal structure of hadrons within the standard model is one of the main challenge of nuclear and particle physics. While the high energy limit of the standard model will soon receive new interest from the expected results at LHC, a full nonperturbative description of relativistic bound state systems in Quantum ChromoDynamics (QCD) is still missing. Many theoretical frameworks already exist and shed some light on these systems, like QCD sum rules, lattice QCD or chiral perturbation theory. All of them have their intrinsic theoretical limitations.

In order to have more physical insights into the internal structure of hadrons, we have thus still to rely on constituent quark models. In the sector of up and down quarks, these models should be relativistic. This is also mandatory if one wants to understand physical observables for which the energy scale can be large, like for instance the electromagnetic and transition form factors at high momentum transfer, or the decay constant of the pion. The interest of a phenomenological analysis of the structure of the pion has been renewed by recent experimental data from the Babar collaboration on the pion transition form factor at high momentum transfer \cite{transition3}. These data (and older ones \cite{transition1,transition2}), as well as known data on the pion electromagnetic form factor \cite{pion1,pion2,pion3,pion4,pion5,pion6,pion7,pion8} and the precise measurement of the pion decay constant \cite{pdg} form a rather large set of data to constrain theoretical models in both the low and high momentum domains.

In the very high momentum transfer limit, factorization theorems enable a simple description of exclusive processes like the electromagnetic or transition form factors of the pion in terms of a distribution amplitude \cite{BL,mikhailov,noguera,bakulev,li,yakovlev}. This distribution amplitude is an (integrated) amplitude which depends only on the longitudinal momentum fraction of the constituent quark. Corrections from the finite transverse momentum of the constituents may however contribute significantly at low and moderate values of the momentum transfer \cite{cao,musatov,raha,jakob,jkr}. Moreover, the full structure of the pion involves two spin (or helicity) components \cite{hww}. These are a-priori of equal importance in this momentum range.

The first requirement in order to build a relativistic dynamical theory of bound state systems  is that it should be invariant 
under the ten generators of the Poincar\'e group. 
These generators include space  time translations (four generators), space rotations (three generators) 
and Lorentz boosts (three generators). 
Following this requirement, three forms of dynamics have been derived by Dirac already in 1949 \cite{Dirac}. 
These are the instant form, the point form and the front form. 

We shall concentrate in this study on the front form. 
In this form of dynamics, the system is defined on slices  
$t^+ = t+z = cte$. 
This form of dynamics is of particular interest
since  the boost operator along the $z$ axis is purely kinematical.  The electromagnetic form factors are thus particularly simple to calculate. However, the plane 
$t^{+}=cte$ is clearly not invariant under all spatial rotations. 
The angular momentum
 operators are therefore dynamical operators. 
In order to treat in a 
transparent way the dependence of these operators on the dynamics, an explicitly
 covariant formulation of light front dynamics (CLFD) has been derivedin Ref.~\cite{karm76}.  
The orientation of the light front plane is here characterized by an 
arbitrary light like four vector $\omega$ with $\omega \, \cd x = cte$.
This approach is a generalization of the 
standard light-front dynamics (LFD) \cite{brodsky}. 
The latter can easily be recovered with
 a special choice of the light-front orientation,  $\omega = (1,0,0,-1)$. 

In the past few years, CLFD has been reviewed~\cite{cdkm} and
 applied to few-body relativistic particle and nuclear physics. 
This formulation is particularly
 appropriate to describe hadrons, and all observables related to them, within the constituent
 quark model. 
The explicit covariance of this formalism is realized by the invariance
 of the light-front plane $\omega \, \cd x = cte$ under any Lorentz transformation. This implies
that $\omega$ is not the same in any reference frame, but varies
according to Lorentz transformations, like the coordinate $x$. It
is not the case in the standard formulation of LFD where $\omega$
is fixed to $\omega=(1,0,0,-1)$ in any reference frame.
Moreover,
 the separation of kinematical and dynamical transformations of the 
state vector provides a definite prescription for constructing bound and scattering
 states of definite angular momentum. 
The dynamical dependence of the wave function becomes a dependence on the position of the light-front defined by $\omega$. 

As we shall see in this study, this explicitly covariant formalism enables 
a very simple analysis of the structure of the two-body bound state. 
The calculation of relativistic corrections, kinematical as well as dynamical, is thus very easy, with a
 clear connection with non-relativistic approaches since it is also three-dimensional. A similar analysis in the heavy quark sector (structure of the $J/\Psi$) has already been done in Ref.~\cite{dugne}.

In order to constrain the phenomenological
structure of hadron wave functions \cite{kroll,belyaev,schlumpf}, one needs to consider  several physical
observables. 
In the case of the pion, this includes the decay constant, the electromagnetic form factor and  the
transition form factor. In our phenomenological study, these observables are calculated in the relativistic impulse approximation.
Since our formalism is fully relativistic and can handle the full structure of the pion wave function - in terms of two spin amplitudes -  it can describe low as well as high momentum scales, with its full kinematical structure in terms of both the longitudinal momentum fraction and the transverse momentum of the constituent quark and antiquark. This is at variance with most of the previous studies which deal with a single distribution amplitude of the pion \cite{BL,mikhailov,noguera,bakulev,li,yakovlev} which may be corrected from transverse momentum contributions \cite{cao,musatov,raha,jakob,jkr}. A first analysis of the electromagnetic form factor of the pion with the full structure of the pion wave function within the standard formulation of LFD can be found in Ref.~\cite{hww}.

The remainder of this paper is organized as follows. 
In section \ref{CLFD}, we present the basic properties of CLFD.
We apply in section \ref{pionwf} our formalism to  the pion wave function, and calculate the physical observables in section \ref{obser}. The numerical results are discussed in section \ref{numerics}.
We summarize our results and  present our conclusions in section \ref{conc}. 

\section{Covariant formulation of light-front dynamics} \label{CLFD}
The description
 of relativistic systems in CLFD has several nice features particularly convenient in the framework of the relativistic constituent quark model. The most important ones are:
\begin{itemize}
\item the formalism does not involve vacuum fluctuation contributions. Therefore, the
 state vector describing the physical bound state contains a definite number of particles, 
as given by Fock state components;
\item the Fock components of the state vector satisfy a three dimensional
equation, and the relativistic wave function has the same interpretation as a probability
 amplitude, like the non-relativistic one;
 \item relativistic wave functions and off-shell amplitudes have a dependence on the
 orientation of the light-front plane which is fully parametrized in terms of the four 
vector $\omega$.  In general, approximate on-shell physical amplitudes  also depend on $\omega$, 
whereas, exact,  on-shell physical amplitudes do not depend on the orientation of the 
light-front plane. This spurious dependence is explicit in CLFD and is therefore under strict theoretical control.
\end{itemize}

\begin{figure}[tb]
\begin{center}
\includegraphics[width=12pc]{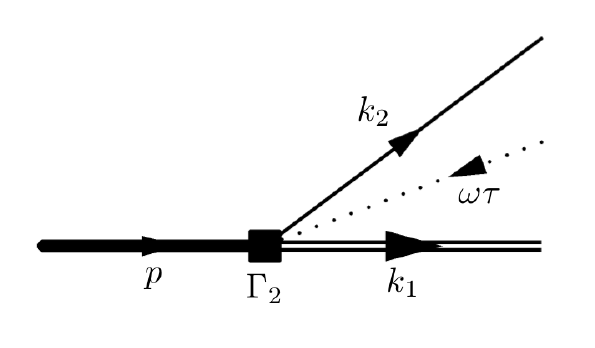}
\end{center}
 \caption{Representation of the two-body wave function in CLFD. The dotted line represents the off-shell energy of the bound state (spurion), while the thick solid line represents the pion. The quark (antiquark) is shown by a thin solid line (double thin line). The vertex function $\Gamma_2$ is defined in Eq.~(\ref{gam2}).}
 \label{gamma2}
\end{figure}

The physical bound state is described by a state vector
 expressed in terms of Fock components. The state vector is  an irreducible
 representation of the Poincar\'e group and is  fully defined by its mass, $M$, its 
four momentum, $p$, its total angular momentum, $J$, and the $z$-axis projection of 
its angular momentum, $\lambda$. The state vector, ${| p, \lambda \rangle}_{\omega}$ of the 
pion of momentum $p$, 
defined on a light-front plane characterized by $\omega$ (with $\omega \cdot x =0$ for 
simplicity), is given in the two-body approximation by \cite{cdkm}
\begin{eqnarray}\label{eq1}
{| p, \lambda \rangle}_{\omega}& =&(2 \pi)^{3/2} \sum_{\sigma_1,\sigma_2}\int \Phi_{
\sigma_{1}\sigma_{2}}^{\lambda}
(k_{1},k_{2},p,\omega \tau)\nonumber \\
&&b_{\sigma_{1}}^{\dagger}({\bf k}_{1})a_{\sigma_{2}}^{\dagger}
({\bf k}_{2})|0
\rangle \delta^{(4)}(k_{1}+k_{2}-p-\omega \tau)  \nonumber \\
&&2(\omega \cd p) {\rm d}
\tau \frac{{\rm d}^{3}
{\bf k}_{1}}
{(2 \pi)^{3/2}\sqrt{2 \varepsilon_{k_1}}} \frac{{\rm d}^{3}{\bf k}_{2}}
{(2 \pi)^{3/2}\sqrt{2 \varepsilon_{k_2}}}  \ ,
\end{eqnarray}
where ${\bf k}_{i}$ is the
 momentum of the quark (or antiquark) $i$, of mass $m$, and $\varepsilon_{k_i}=\sqrt{{\bf k}_{i}^2+m^2}$.  The creation operators for the antiquark and quark are denoted by $b^{\dagger}$ and $a^{\dagger}$ respectively; $\lambda$ is the projection of the total angular momentum 
of the system on the $z$ axis in the rest frame and $\sigma_{i}$
 is the spin projections of the particle $i$ in the corresponding rest system. 
From the delta  function, $\delta^{(4)}(k_{1}+k_{2}-p-\omega \tau)$, ensuring momentum conservation, one gets
\begin{equation}\label{eq2}
\mathcal{P} \equiv p+ \omega \tau = k_{1}+k_{2}\ .
\end{equation}
This peculiar momentum conservation law arises directly from the invariance of the reference system under translations along the light-front time \cite{cdkm}.
It is convenient to represent  this conservation law in a systematic way. To do that, 
we shall represent on any diagram 
the four-vector $\omega \tau$ by a dotted line (the so-called 
spurion line, see~\cite{cdkm} for more 
details), with an orientation opposite to the quark and antiquark momenta.  The two-body wave function $\Phi$ will thus be represented by the diagram of Fig.~\ref{gamma2}. We emphasize that the bound state wave function is always an off-energy 
shell object ($\tau \neq  0$  due to binding energy) and depends therefore on the light-front
 orientation. The parameter $\tau$ is
 entirely determined by the on-mass shell condition for the individual constituents, and the conservation law (\ref{eq2}).
The state vector
 is normalized according to 
\begin{equation}\label{eq30}
\langle p^{\prime},\lambda^{\prime} | p, \lambda \rangle = 2 \varepsilon_p \delta^{(3)}(\bf p -
\bf p^{\prime}) \delta^{\lambda^{\prime} \lambda}\ .
\end{equation} 

The two-body wave function $\Phi(k_1,k_2,p,\omega \tau)$ written
 in Eq.~(\ref{eq1}) can be parametrized in terms of various sets of kinematical variables. In 
order to make a close connection to the non-relativistic case, it is more 
convenient to introduce the following variables~\cite{cdkm} defined by
\begin{equation}\label{eq3}                                                     
{\bf k}=L^{-1}({ \bf \cal P}){\bf k}_1 = {\bf k}_1 -\frac{{\bf \cal P}}{\sqrt{{\cal P}^2}}
\left[k_{10}-
 \frac{{\bf k}_1\cd {\bf {\cal  P}}}{\sqrt{{\cal P}^2}+{\cal P}_0}\right]\ ,     
\end{equation}
\begin{equation}\label{eq4}                                                    
{\bf n}=
\frac{L^{-1}({\cal P}){\bf \omega}}{\vert L^{-1}({\cal P}){\bf \omega}\vert} 
=\sqrt{{\cal P}^2}\frac{L^{-1}({\cal P}){\bf \omega}}{\omega\cd p}\ ,
\end{equation}    
where  $L^{-1}({\cal P})$ is the (inverse) Lorentz boost of momentum ${\cal P}$. 
The momentum $\bf{k}$ corresponds, in the frame where 
${\bf k}_1+{\bf k}_2={\bf 0}$, to the usual relative momentum between the two 
particles. The unit vector $\bf{n}$ corresponds, in this frame, to the spatial direction of
 $\bf{\omega}$. Note that this choice of variable does not assume that we restrict ourselves
 to this particular frame. 

The second set of variables which we shall also use in the following is the
 usual light-front set of coordinates $(x,\bf{R}_{\perp})$, which is defined by 
\begin{eqnarray}\label{eq5}
x&=& \frac{\omega \cd k_{1} }{ \omega \cd p} \nonumber \ ,\\
R_{1}&=&k_{1}-xp \ ,\nonumber 
\end{eqnarray}
and where $R_1$ is decomposed into its spatial components parallel and perpendicular to the
 direction of the 
light-front,   $ R_{1}=(R_{0},{\bf R}_{\perp}, {\bf R}_{\|})$. We have by definition $R_1
 \cd \omega=0$,
 and thus $R_1^2=-{\bf R}
_{\perp}^2$. In the reference frame where ${\bf p_\perp}=0$,  ${\bf R}_{\perp}$ is identical to the usual transverse momentum ${\bf k}_{\perp}$.
The relations between these two sets of variables are given by
\begin{eqnarray}\label{eq6}
{\bf R}_{\perp}^{2}& = {\bf k}^{2} - ({\bf n} \cd {\bf k})^2 \ , \nonumber \\
x & = \frac{1}{2} \bigg[ 1 - \frac{({\bf n} \cd {\bf k})}{\epsilon_{k}}\bigg]\ .
\end{eqnarray}
The inverse relations read:
\begin{eqnarray}\label{eq7}
{\bf k}^{2}& = &\frac{{\bf R}_{\perp}^{2} + m^2}{4x(1-x)} - m^2 \ ,  \nonumber \\
{\bf n} \cd {\bf k}& = &\bigg[ \frac{{\bf R}_{\perp}^{2}+m^2}{x(1-x)} \bigg]^{1/2}
 \bigg(\frac{1}{2}
 - x\bigg)\ .
\end{eqnarray}
Note that ${\bf k}^{2}$ and ${\bf n} \cd {\bf k}$ are invariant under any rotation and 
 Lorentz boost \cite{cdkm}, like $x$ and ${\bf R}_\perp^2$. In the non relativistic limit, ${\bf n} \equiv {\bf n}/c \to 0$ and therefore $x\to 1/2$ and ${\bf n}.{\bf k} \to 0$.

\section{The pion wave function} \label{pionwf}
\subsection{Structure of the bound state}
The  covariance of our approach allows to write down explicitly the general spin
 structure of the two-body bound state. For a pseudoscalar particle of momentum $p$, composed of an 
 antiquark and a quark of equal masses $m$ and of momenta $k_1$ and $k_2$, respectively, it takes the form
\begin{equation}\label{eq11}                                                    
\Phi_{\sigma_1 \sigma_2}^{\lambda=0}=\frac{1}{\sqrt{2}}\bar{u}_{\sigma_2}(k_2)\left(A_1\frac{1}{m}+ A_2
\frac{\sla{\omega}}{\omega\cd p}\right) \gamma_5\  v_{\sigma_1}(k_1)\ ,
\end{equation}
where  $v(k_{1})$ and  $u(k_{2})$  are the usual Dirac spinors, and $A_1$ and
 $A_2$ are the two scalar components of the pion wave function. For simplicity, we shall also call wave functions these two spin components. They depend on two scalar variables, which we shall choose as $(x,{\bf R}_{\perp}^{2})$. We do not show for simplicity the standard isospin and color components of the pion wave function in Eq.~(\ref{eq11}). 
The representation of this wave function in terms of the
 variables $\bf{k}$ and $\bf{n}$ is given by
\begin{equation}\label{eq12}
\Phi_{\sigma_1 \sigma_2}^0=\frac{1}{\sqrt{2}}w_{\sigma_2}^t\left(g_1+\frac{i\bf{\sigma}\cd [\bf{n}\times 
\bf{k}]}{k}g_2\right)w_{\sigma_1}\ ,
\end{equation} 
where $w_{i}$ are Pauli spinors and $g_{1,2}$ are the two scalar  components of the pion wave function in this representation.  They depend also on two scalar variables, which we shall choose as $({\bf k}^2, {\bf k}.{\bf n})$. One can easily express $A_{1,2}$ in terms of $g_{1,2}$. We get~\footnote{One uses here 
the standard definition of $\gamma^5$
 with positive sign for its matrix elements, contrarily to~\cite{cdkm} where $\gamma^5$ 
has an opposite sign.}
\begin{eqnarray}\label{eq13}
g_1 &= &\frac{2 \varepsilon_k}{m} A_1+\frac{m}{\varepsilon_k} A_2 \ ,\\
g_2 &=&-\frac{k}{\varepsilon_k} A_2\ .
\end{eqnarray}
We would like to stress that the decomposition (\ref{eq11}) is a very general one for a spin
 zero particle composed of two spin 1/2 constituents. In the non-relativistic limit, 
 the component $g_1({\bf k}^2, {\bf k}.{\bf n})$ only survives and depends on a single scalar variable ${\bf k}^2$. 
In our phenomenological analysis, we shall therefore start from a  non-relativistic 
component, $g_1^0$, given by a simple parametrization. We shall use in the following either a gaussian wave function given by
\begin{equation}\label{eq15}
g_1^0({\bf k}^2)=\alpha \  {\rm exp}(-\beta \ \bf{k}^2)\ , \\
\end{equation}
or a power-law wave function written as
\begin{equation}
\label{eq15b}
g_1^0({\bf k}^2)=\frac{\alpha} {(1+ \beta \ \bf{k}^2)^\gamma}\ , \\
\end{equation}
where $\beta$ is a parameter to be determined from experimental data, while $\alpha$ will be fixed from the normalization condition. The power $\gamma$ will be chosen equal to $2$. The
 relativistic component $A_2$, as well as dynamical relativistic corrections to $A_1$, will 
be calculated from radiative corrections, as explained in the next subsection. The choice (\ref{eq15}) is equivalent to the Brodsky-Huang-Lepage parametrization \cite{BHL}.

The normalization
 condition writes ~\cite{cdkm}
\begin{equation}\label{eq31}
1 = \sum_{\sigma_{1} \sigma_{2}} \int dD \ \Phi^\lambda_{\sigma_{1} \sigma_{2}}
\Phi_{\sigma_{1}\sigma_{2}}^{\lambda \star} \ ,
\end{equation}
where $dD$ is an invariant phase space element which can take the following forms, depending on the kinematical variables which are used
\begin{equation}\label{eq32}
dD=\frac{1}{(2\pi)^3}\frac{{\rm d}^3{\bf k}_1}{(1-x)2\varepsilon_{k_1}}
=\frac{1}{(2\pi)^3}\frac{{\rm d}^3{\bf k}}{2 \varepsilon_k }
=\frac{1}{(2\pi)^3}\frac{{\rm d}^2{\bf R}_{\perp}{\rm d} x}{2x(1-x)}\ .
\end{equation}
With the pion wave function written in Eq.~(\ref{eq11}), the  
 normalization condition writes~\cite{cdkm}
\begin{eqnarray}\label{eq33}
1= \frac{1}{(2\pi)^3} \int &&\frac{{\rm d}^2{\bf R}_{\perp}{\rm d} x}{2x(1-x)}\ 
\Biggr[  \frac{{\bf R}_{\perp}^{2}+m^2}{m^{2}x(1-x)}  A_{1}^{2}  \nonumber \\
&& + 4 A_{1}A_{2}  +  4x(1-x) A_{2}^{2} \Biggl]  \ .
\end{eqnarray}

\subsection{Radiative corrections to the wave function} \label{radia}
In a traditional non-relativistic study of the pion  wave function in the spirit of the 
constituent quark model, one may  start directly from a simple 
parametrization of the component $g_1^0$, as given for instance in Eqs.~(\ref{eq15},\ref{eq15b}). However, it is 
necessary to correct this wave function in some way  in order to incorporate in a full relativistic framework the 
high momentum tail given by the one-gluon exchange mechanism. We
 shall achieve this using  perturbation theory, starting from the zeroth
 order wave function $g_1^0$. 
 
\begin{figure*}[tb]
\begin{center}
\includegraphics[width=40pc]{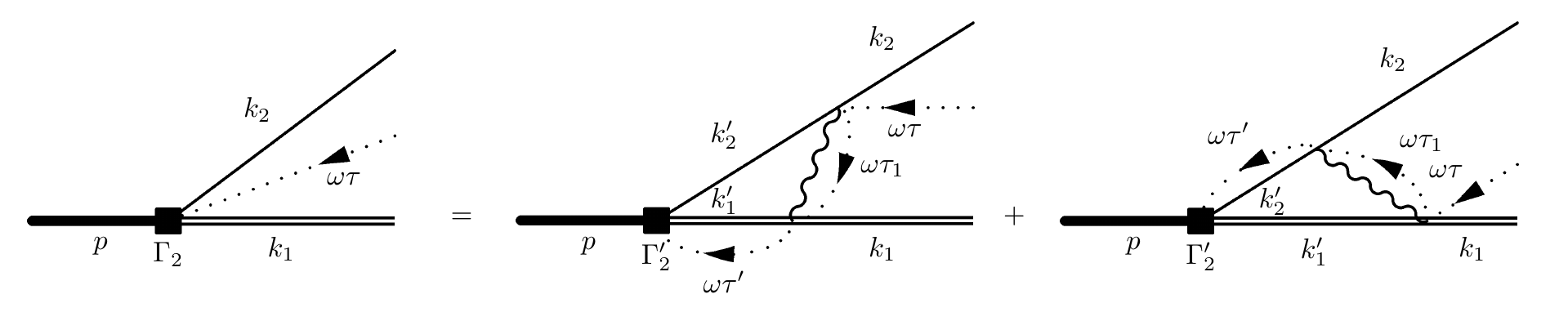}
\end{center}
 \caption{Calculation of radiative corrections to the two-body wave function. In this figure, and in all subsequent figures, all particle lines are oriented from the left to the right.}
 \label{equation}
\end{figure*}

 The (eigenvalue) equation we start from to calculate the bound state wave function is represented schematically in Fig.~\ref{equation}. According to the diagrammatic rules of CLFD, this equation writes, in the case of spin $1/2$ particles \cite{cdkm}
\begin{eqnarray}\label{eq16} 
&&\bar u(k_2) \Gamma_2 v(k_1)=
\int \frac{{\rm d}^3{\bf k_1}}{ 2\varepsilon_{k_1}(2\pi)^3}\frac{{\rm d} \tau'}{\tau' -i \epsilon}  \delta (k_2^{\prime 2}-m^2)\Theta(\omega \cd k'_2)\nonumber \\
&&\times \bar u(k_2) \left[ \gamma_\mu (\sla k_2^\prime +m)
\Gamma_2^\prime
(m-\sla k_1^\prime)
\gamma_\nu \right] K^{\mu\nu} v(k_1)\  .
\end{eqnarray}
It is written in terms of the two-body vertex function $\Gamma_2$ defined by \cite{kms_08}
\begin{equation} \label{gam2}
\bar{u}(k_2) \Gamma_2  v(k_1) \equiv (s-M^2_\pi) \Phi\ ,
\end{equation}
with
\begin{eqnarray}
s=\frac{{\bf R}_{\perp}^2+m^2}{x}+\frac{{\bf R}_{\perp}^2+m^2}{(1-x)}\ .
\label{eq17}
\end{eqnarray}
The mass of the pion is denoted by $M_\pi$. We shall define for simplicity
\begin{equation} \label{vartheta}
{\cal O}\equiv \frac{\Gamma_2}{(s-M^2_\pi)}=\frac{1}{\sqrt{2}}\left(A_1\frac{1}{m}+ A_2
\frac{\sla{\omega}}{\omega\cd p}\right)  \gamma_5 \ ,
\end{equation}
and similarly for $\cal O '$ in terms of prime quantities. The components $A_{1,2}$ depend on $(x,{\bf R}_{\perp}^2)$, while $A'_{1,2}$ depend on $(x',{\bf R'}_{\perp}^2)$.

The  kernel, 
$K^{\mu\nu}$, including the appropriate color factor, can be 
written as $K^{\mu\nu}=-g^{\mu\nu} \frac{4}{3} g^2 \mathcal{K}$ in the Feynman gauge, with
\begin{eqnarray}\label{eq19} 
\mathcal{K} &= &\int \theta \left[\omega   \cd    (k_{1} -  k_{1}^{\prime})\right]
\delta \left[(k_{1} - k_{1}^{\prime} + \omega \tau_1 -  
\omega \tau)^2\right] \frac{d \tau_1}{\tau_1-i \epsilon} \nonumber \\
&+ &\int \theta \left[\omega   \cd    (k_{1}^{\prime}  -  k_{1})\right] \delta \left[(k_{1}^{\prime} 
 -  k_{1}   + 
\omega \tau_1 -  \omega \tau^{\prime})^2 \right] \frac{d \tau_1}{\tau_1-i\epsilon} \ ,\nonumber \\
\end{eqnarray}
where $\tau$  and $\tau^{\prime}$ are defined by
\begin{equation}\label{eq21} 
\tau = \frac{s-M_\pi^{2}}{2 \; \omega \cd p}\ , \;\;\; {\rm and}
 \;\;\;\tau^{\prime} = \frac{s^{\prime}-M_\pi^{2}}{2 \; \omega \cd p}\ .
\end{equation}
After integration over $\tau_1$, we have
\begin{eqnarray}\label{eq20} 
\mathcal{K}& = &\frac{\theta \left[\omega   \cd    (k_{1} -  k_{1}^{\prime})\right]}{ -(k_{1}
 -  
k_{1}^{\prime})^2+ 2 \tau \omega \cd (k_{1} -  k_{1}^{\prime})}  \nonumber \\
&+ &\frac{\theta \left[\omega   \cd    (k_{1}^{\prime} -  k_{1})\right]}{ -(k_{1}^{\prime} - 
 k_{1})^2
+ 2 \tau^{\prime} \omega \cd (k_{1}^{\prime} -  k_{1})}\ .
\end{eqnarray}
Using the scalar products calculated in Appendix D of Ref.~\cite{cdkm}, one gets the following final expression for
 $\mathcal{K}$
\begin{equation}\label{eq24}
\mathcal{K}=
\frac{x^{\prime}(x-1)\theta(x-x^{\prime}) }{\mathcal{K}_{>}}
+\frac{x(x^{\prime}-1)\theta(x^{\prime}-x) }{\mathcal{K}_{<}}\ ,
\end{equation}
where,
\begin{eqnarray}\label{eq22} 
\mathcal{K}_{>}&= &m^2(x-x')(x-x'-1) \\
&&+{\bf R_\perp}^2x'(x'-1)+{\bf R'_\perp}^2x(x-1)\nonumber \\
&&-M_\pi^2x'(x-1)(x-x')-2x'(x-1){\bf R_\perp}.{\bf R'_\perp} \ , \nonumber
\end{eqnarray}
and
\begin{eqnarray}\label{eq23} 
\mathcal{K}_{<}=&= &m^2(x-x')(x-x'-1) \\
&&+{\bf R_\perp}^2x'(x'-1)+{\bf R'_\perp}^2x(x-1)\nonumber \\
&&+M_\pi^2x(x'-1)(x-x')-2x(x'-1){\bf R_\perp}.{\bf R'_\perp} \ . \nonumber
\end{eqnarray}
The quark gluon coupling constant is denoted by $g$, with $g^2=4\pi \alpha_s$. In order to incorporate the correct short range properties of the quark-antiquark interaction from asymptotic freedom, we shall consider in the following a running coupling constant $\alpha_s(K^2)$, where $K^2$ is the off-shell momentum squared of the gluon. It is given by $K^2=1/\mathcal{K}$. We choose a simple parametrization which gives, in the large $K^2$ limit, the known behavior given by perturbative QCD. We take
\begin{equation} \label{alphas}
\alpha_s(K^2)=\frac{\alpha_s^0}{1+\frac{11-\frac{2}{3} n_f}{4 \pi} \alpha_s^0 \mbox{Log}\left[ \frac{\vert K^2 \vert+\Lambda^2_{QCD}}{\Lambda^2_{QCD}}\right]} \ .
\end{equation}
At small $K^2$, it is given by the parameter $\alpha_s^0$ which should be of the order of $1$. We choose $n_f=2$ and $\Lambda_{QCD}=220$ MeV.

In order to extract the two components $A_{1,2}$, one should proceed as follows. We
 first multiply both sides of Eq.~(\ref{eq16}) by $u(k_2)$ on the left and $\bar v(k_1)$ 
on the right, and sum over polarization states. We then multiply both sides successively 
 by $\gamma_{5}$ and $ \omega\!\!\!\!/ \gamma_{5}$, and take the trace. We end up with
 the following system of two equations for the two unknowns $A_{1,2}$
\begin{eqnarray}\label{eq25}  
{\rm Tr} \Bigl[ \gamma_5(\sla k_2+m) {\cal O} (\sla k_1-m) \Bigr]= \frac{1}{(s-M_\pi^2) (2 \pi)^3} \nonumber \\
\times \int 
{\rm Tr} \Bigl[ \gamma_5(\sla k_2+m)A'_{\mu \nu} (\sla k_1-m) \Bigr]K^{\mu \nu}
\frac{{\rm d}^{2} {\bf{R}}
_{\perp}^{\prime} {\rm d}x^{\prime}}{2x^{\prime}(1-x^{\prime})}\ , \nonumber \\
\end{eqnarray}
and
\begin{eqnarray}\label{eq26} 
{\rm Tr} \Bigl[ \sla \omega \gamma_5(\sla k_2+m) {\cal O} (\sla k_1-m)
 \Bigr]= \frac{1}{(s-M_\pi^2) (2 \pi)^3} \nonumber \\
\times \int 
{\rm Tr} \Bigl[ \sla \omega \gamma_5(\sla k_2+m)A'_{\mu \nu} (\sla k_1-m)
 \Bigr]K^{\mu \nu} \frac{{\rm d}^{2} {\bf{R}}_{\perp}^{\prime} {\rm d}x^{\prime}}
{2x^{\prime}(1-x^{\prime})}\ ,\nonumber \\
\end{eqnarray}
with $A'_{\mu \nu}$ defined by
\begin{equation}\label{eq27} 
A'_{\mu \nu} = \gamma_\mu (\sla k_2^{\prime} +m) {\cal O}^{\prime}(m-\sla 
 k_1^{\prime}) \gamma_{\nu}\ .
\end{equation}
In perturbation theory, we shall start from a non-relativistic pion wave function given by
\begin{equation} \label{O0}
{\cal O}^0=\frac{1}{\sqrt{2}} \frac{A_1^0}{m} \gamma_5\ ,
\end{equation}
where $A_1^0$ is calculated  from (\ref{eq15}) or (\ref{eq15b}) with $g_1=g_1^0$ and $g_2=0$. The correction to the wave function coming from one gluon exchange is denoted by $\delta {\cal O}$ and given by
\begin{equation} \label{do}
\delta {\cal O}=\frac{1}{\sqrt{2}}\left(\delta A_1\frac{1}{m}+ \delta A_2
\frac{\sla{\omega}}{\omega\cd p}\right)  \gamma_5\ .
\end{equation}
It is calculated from Eqs.~(\ref{eq25},\ref{eq26}) with the replacement ${\cal O} \to \delta {\cal O}$ in the l.h.-s. and ${\cal O}' \to {\cal O}'^0$ in the r.h.-s.. The total wave function is then given by
\begin{equation}
{\cal O} = {\cal O}^0 + \delta {\cal O} \ ,
\end{equation}
with
\begin{eqnarray}\label{eq29} 
\delta A_{1}(x,{\bf R}_{\perp}^2) &= &\frac{1}{(s-M^2) 2 \pi^2}\int \frac{{\rm d}^{2} {\bf{R}}_{\perp}^{\prime} {\rm d} x^{\prime}}{2x^{\prime}(1-x^{\prime})}  \mathcal{K} A_1^{\prime 0} \alpha_s(K^2) \nonumber \\
&\times & \frac{m^2(2x'^2-2x'+1)+{\bf R'}_{\perp}^2}{x'(1-x')}\ ,
\end{eqnarray}
\begin{eqnarray}\label{eq29b}
&&\delta A_{2}(x,{\bf R}_{\perp}^2) = \frac{1}{(s-M^2) 2 \pi^2}\int \frac{{\rm d}^{2} {\bf{R}}_{\perp}^{\prime} {\rm d} x^{\prime}}{2x^{\prime}(1-x^{\prime})} \mathcal{K} A_1^{\prime 0}  \alpha_s(K^2) \nonumber \\
&&\frac{m^2(x-x')(x+x'-1)+{\bf R}_{\perp}^2 x'(1-x')-{\bf R'}_{\perp}^2 x(1-x)}{x(1-x)x'(1-x')} \ .\nonumber\\
\end{eqnarray}
It is instructive to exhibit the behavior of these components at very high transverse momentum $\vert {\bf{R}}_{\perp} \vert $. From Eqs.~(\ref{eq29},\ref{eq29b}), and using (\ref{eq24}), it is easy to see that $\delta A_1$ is, in the absence of the running coupling constant,  of the order of $1/{\bf{R}}_{\perp}^4$ while $\delta A_2$ is of the order of $1/{\bf{R}}_{\perp}^2$. The running of the coupling constant adds a factor $1/\mbox{Log}\left[\frac{{\bf{R}}_{\perp}^2}{\Lambda_{QCD}^2}\right]$. At high transverse momentum, the relativistic component $A_2=\delta A_2$ thus dominates. More precisely, we have, in this limit
\begin{eqnarray} \label{A2inf}
A_2 \to A_2^\infty& =& \frac{2}{3 \pi^2} \frac{1}{{\bf{R}}_{\perp}^2 \mbox{Log}\left[\frac{{\bf{R}}_{\perp}^2}{\Lambda_{QCD}^2}\right]} \int_0^1 \frac{{\rm d} x^{\prime}}{2x^{\prime}(1-x^{\prime})} \frac{A_1^{'0}}{\bar K} \nonumber \\
&\equiv &  \frac{1}{{\bf{R}}_{\perp}^2 \mbox{Log}\left[\frac{{\bf{R}}_{\perp}^2}{\Lambda_{QCD}^2}\right]}  \bar A_2^\infty(x)\ ,
\end{eqnarray}
where $\bar K=x'/x$ for $x<x'$ and $\bar K=(1-x')/(1-x)$ for $x>x'$.

\section{Physical observables}  \label{obser}

\subsection{Decay constant}
\begin{figure}[b]
\begin{center}
\includegraphics[width=12pc]{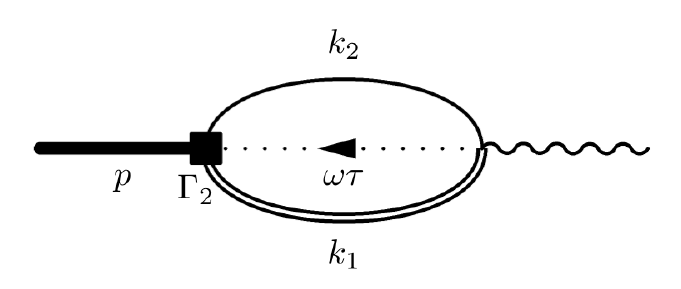}
\end{center}
 \caption{Decay amplitude of the pion.}
 \label{decay}
\end{figure}

The pseudoscalar decay amplitude is given by the diagram  in Fig.~\ref{decay}. 
 According to the usual definition, the decay amplitude  is  ${\Gamma}_{\mu}= \langle0 |
 J_{\mu}^5 | \pi \rangle$ where $J_{\mu}^5$
 is the axial current. Since our formulation is explicitly covariant, 
we can decompose ${\Gamma}_{\mu}$
 in terms of all four-vectors available in our system, i.e. the incoming meson momentum 
$p$ and the arbitrary position, $ \omega$, of the light-front. We have 
therefore
\begin{equation}\label{eq34}
{\Gamma}_{\mu}= F\ p_{\mu} + B\  \omega_{\mu}\ ,
\end{equation}
where  $F$ is the physical pion decay constant. 
In an exact calculation,
$B$ should be  zero, while it is a priori non zero in any approximate calculation. It is a non physical, spurious, contribution which should be extracted from the full amplitude $\Gamma_\mu$.
Since $\omega^2=0$, the physical part of the pion decay constant  can  easily be obtained from
\begin{equation}\label{eq35}
 F = \frac{\Gamma \cd \  \omega}{\omega \cd p}\ .  
\end{equation}
Using the diagrammatic rules of CLFD \cite{cdkm}, we can calculate $\Gamma_\mu$ from the graph indicated
 in Fig.~\ref{decay}. 
One gets, including color factors,
\begin{eqnarray}\label{eq36}
{\Gamma}_{\mu}&=&\sqrt{3} \int \frac{{\rm d}^3{\bf k_1}}{2\varepsilon_{k_1}}\frac{{\rm d} \tau}{\tau -i \epsilon}  \delta (k_2^{ 2}-m^2)\Theta(\omega \cd k_2) \nonumber \\
 &&\times {\rm Tr} \left[ -\overline{ \gamma_{\mu}\gamma_{5}}(\sla{k_{2}} +
m) \Gamma_2
(m - \sla{k_{1}}) \right] \ ,
\end{eqnarray}
with $\Gamma_2$ defined in (\ref{vartheta}) and where the notation $\overline O$ means as usual
\begin{equation}\label{eq37}
 \overline{O} = \gamma^{0} O^{\dagger} \gamma^{0}\ .
\end{equation}
After reduction of the scalar products, the decay constant is thus given by
\begin{equation}\label{eq38}
F = \frac{2 \sqrt{6}}{(2\pi)^3} \int \frac{{\rm d}^{2} {\bf{R}}_{\perp} {\rm d} x}{2x(1-x)} \left[ 
A_{1}+2x(1-x)A_{2} \right] \ .
\end{equation} 

One can immediately notice that the pion decay constant given by (\ref{eq38}) is divergent with the asymptotic relativistic component $A_2^\infty$ given by (\ref{A2inf}). It diverges like $\mbox {Log Log} \  {\bf{R}}_{\perp}^2/\Lambda_{QCD}^2$. This divergence is extremely soft.  It is in fact the expression of the well known divergence of radiative corrections in the process $q \bar q \to \gamma$. To get the physical contribution, we just subtract the minimal contribution arising when the integral on $\vert {{\bf R}}_{\perp}\vert $ is cut-off to $\Lambda_C$. The physical pion decay constant is thus
\begin{equation}
F^{phys} = F - F^{\infty}\ ,
\end{equation}
where
\begin{equation}
F^\infty=\frac{2 \sqrt{6}}{4\pi^2} \frac{1}{\mbox{Log Log} \ \left[ \frac{\Lambda_C}{\Lambda_{QCD}}\right] }\int_0^1 \frac{dx}{2x(1-x)} \bar A_2^\infty(x) \ ,
\end{equation}
in the limit where $\Lambda_C$ is very large, with $\bar A_2^\infty$ defined in Eq.~(\ref{A2inf}).

\subsection{Electromagnetic form factor}
The electromagnetic form factor is one of the most useful observable which can   be used to
probe the internal structure of a bound state. Moreover,  from the electromagnetic
 form factor at very low momentum transfer, it is possible to determine the charge radius of the composite particle. This physical observable is therefore very powerful in order to
 constrain the phenomenological structure of the wave function both in the low and high momentum domains. 
\begin{figure}[b]
\begin{center}
\includegraphics[width=15pc]{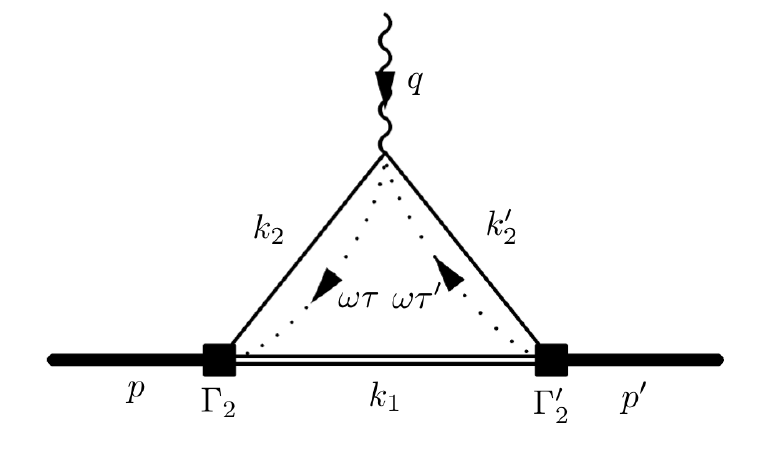}
\end{center}
 \caption{Pion electromagnetic form factor  in the impulse approximation. A similar contribution where the photon couples to the antiquark is not shown for simplicity.}
 \label{elm}
\end{figure}

 In the impulse approximation, the electromagnetic form factor  is shown in Fig.~\ref{elm}. 
In CLFD, the general physical electromagnetic amplitude of a spinless system  can be 
decomposed as~\cite{cdkm}
\begin{equation}\label{eq39}
J ^{\rho}= \langle \pi(p') | e_q\  \bar{q} \gamma^{\rho} q | \pi(p) \rangle
= e_\pi (p+p^{\prime})^{\rho}\ F_{\pi}(Q^{2}) + \frac{\omega^{\rho}}{\omega \cd p}B_1(Q^{2})\ ,
\end{equation}
where $e_q$ is the charge of the quark, while $e_\pi$ is the charge of the pion. The physical form 
factor is denoted by $F_{\pi}(Q^{2})$. In any exact calculation, $B_1(Q^{2})$
 should be zero. We choose for convenience a reference frame where $\omega \cd q=0$, with $q=p'-p$. This implies automatically
 that the form factors $F_{\pi}(Q^{2})$ and $B_1(Q^{2})$ depend on $Q^{2}=-q^{2}$ only, 
since from homogeneity arguments their dependence on $\omega$ is of the form $\omega  \cdot  p / \omega  
\cdot  p^\prime \equiv 1$. The physical electromagnetic form factor $F_{\pi}(Q^{2})$ can
 be simply extracted from $J^\rho$ by contracting both sides of  Eq.~(\ref{eq39}) with $\omega_{\rho}$. One
 thus has
\begin{equation}\label{eq40}
F_{\pi}(Q^{2})= \frac{J \cd \omega }{2 \; \omega \cd p}\ .
\end{equation}
By using the diagrammatic rules of CLFD, we can write down the
 electromagnetic amplitude corresponding to Fig.~\ref{elm} where the photon interacts
 with the quark. Assuming it is pointlike, one  obtains:
\begin{eqnarray}\label{eq41}
F_{\pi}^{\gamma q}(Q^{2}) &= & e_{q} \int \frac{{\rm d}^3{\bf k_1}}{2\varepsilon_{k_1}}\frac{{\rm d} \tau}{\tau -i \epsilon}  \delta (k_2^{ 2}-m^2)\Theta(\omega \cd k_2) \nonumber \\
&\times&\frac{{\rm d} \tau'}{\tau' -i \epsilon}  \delta (k_2^{ \prime 2}-m^2)\Theta(\omega \cd k'_2) \nonumber \\
&\times&  \ {\rm Tr} \Biggr[  {- \bar {\Gamma_2}}^{\prime}
 (\sla k_{2}
^{\prime}+m) \frac{\sla \omega}{2 \omega \cd p}  (\sla k_{2}+ m) 
\Gamma_2(m-\sla k_{1})  \Biggl] \ , \nonumber \\
\end{eqnarray}
where $\Gamma_2$  is given in 
Eq.~(\ref{vartheta}), and similarly for $\Gamma_2^{\prime}$ with prime quantities. After 
calculation of the trace,  one gets
\begin{eqnarray}\label{eq42}
F_{\pi}^{\gamma q}(Q^{2})&= & \frac{e_q}{(2\pi)^3} \int \frac{{\rm d}^{2} {\bf{R}}_{\perp} {\rm d} x}{2x(1-x)}  \Biggr[ \frac{m^2
+{\bf R}_{\perp}^{2}-x{\bf R}_{\perp} \cd {\bf \Delta}} {x(1-x) m^{2}}A_{1}
A_{1}^{\prime} \nonumber \\
&+&  2 (A_{1}A_{2}^{\prime} + 
A_{1}^{\prime}A_{2})  + 4x(1-x)A_{2}A_{2}^{\prime}
 \Biggl] \ .
\end{eqnarray}
The wave functions $A_{1,2}^\prime$  
depend on 
$(x^\prime, {\bf R'}_{\perp}^2)$, with $x^\prime=x$ in the impulse approximation.
If we define the four momentum transfer $q$ by $q=(q_{0},{\bf \Delta},{\bf q}_{\|})$, 
with ${\bf \Delta} \cd {\bf \omega} =0$ and ${\bf q}_{\|}$ parallel to ${\bf \omega}$, 
 we have $Q^{2} = 
-q^{2} \equiv {\bf \Delta}^{2}$, and thus ${\bf R}_{\perp}^{\prime}
={\bf R}_{\perp}-x{\bf \Delta}$. 
The contribution from the coupling of the photon to the antiquark can be deduced from~(\ref{eq42}) 
by the interchange $x \iff (1-x)$, $ {\bf R}_{\perp} \iff -{\bf R}_{\perp}$ and an overall change of sign.

One thus obtains the full contribution to the electromagnetic form factor of the pion
\begin{equation}\label{eq43}
F_{\pi}(Q^{2})= F_{\pi}^{\gamma q}(Q^{2})+F_{\pi}^{\gamma \bar{q}}(Q^{2})\ .
\end{equation}
Note that this form factor, in the impulse approximation,  is completely finite since it does not correspond to any radiative corrections at the $\gamma q$ vertex.
The charge radius of the pion, $\langle r_{\pi}^{2} \rangle^{1/2} $, can be extracted from $F_\pi(Q^2)$ according to
\begin{equation}\label{eq44}
\langle r_{\pi}^{2} \rangle = -6 \frac{{\rm d}}{{\rm d} Q^{2}} F_{\pi}(Q^{2}) \Big 
|_{Q^{2}=0}\ .
\end{equation}

In the very high $Q^2$ limit, it is now well accepted that the pion form factor behaves like 
$F_{\pi}(Q^2) \sim 1/Q^2$ (up to logarithmic corrections). This asymptotic behavior  is fully determined by the one gluon exchange mechanism. This
 mechanism can either be considered explicitly in the hard scattering amplitude \cite{BL}, or 
incorporated in the relativistic wave function of the meson. In the spirit of the relativistic
constituent quark model,  we adopt here the 
second strategy since it permits to investigate in a unique framework both low and high momentum scales. At asymptotically large $Q^2$, the form factor is dominated by the 
contribution from the relativistic $A'_2$ component in Eq.~(\ref{eq42}) \cite{cdkm} calculated at ${\bf R}_\perp^{\prime 2} \sim {\bf \Delta}^2$. We recover here naturally  the asymptotic behavior of the pion electromagnetic form factor.

\subsection{Transition form factor}
\begin{figure}[b]
\begin{center}
\includegraphics[width=15pc]{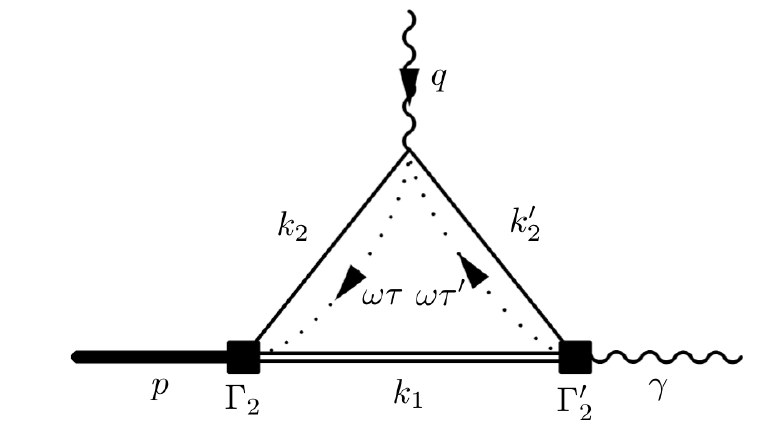}
\end{center}
 \caption{Pion transition form factor in the impulse approximation. A similar contribution where the virtual photon, denoted by $\gamma*$, couples to the antiquark is not shown for simplicity.}
 \label{transition}
\end{figure}

The quantum numbers of the $\pi$ transition amplitude, $\pi \to \gamma^* \gamma$, 
are similar to  the ones of the deuteron electrodisintegration amplitude near threshold, as
 detailed in Ref.~\cite{cdkm}. The exact physical amplitude, $\Gamma_\rho$,  writes therefore
\begin{equation}\label{eq45}
\Gamma_\rho = F_{\mu \rho}\  e^{\mu *} \  ,
\end{equation}
with the amplitude $F_{\mu \rho}$ given by
\begin{equation}\label{eq46}
F_{\mu \rho}= \frac{1}{2} \varepsilon_{\rho \mu \nu \lambda} q^{\nu} P^{\lambda}F_{\pi \gamma}\ ,
\end{equation}
and where  $e^\mu$ is the polarization vector of the final (on-shell) photon. The momenta $P$ and $q$ are defined by 
$P=p+p^{\prime}$ and $q=p^{\prime}-p$ with the kinematics indicated on Fig.~\ref{transition}. In any approximate calculation, the 
amplitude $F_{\mu \rho}$ depends on $\omega$. It should be decomposed 
in terms of all possible tensor structures compatible with the quantum numbers of the
 transition, as we did above for the  decay constant and the electromagnetic form factor. 
 One thus has~\cite{cdkm}
\begin{eqnarray}\label{eq47}
F_{\mu \rho}&= &\frac{1}{2}\varepsilon_{\rho \mu \nu \gamma}q^{\nu}P^{\gamma}F_{\pi \gamma}+
 \varepsilon_{\rho \mu \nu
 \gamma}q^{\nu}\omega^{\gamma}B_{1}+ \varepsilon_{\rho \mu \nu \gamma}p^{\nu}\omega^{\gamma}B_{2} \nonumber \\
 &+&(V_{\mu}q_{\rho}+V_{\rho}q_{\mu})B_{3}+(V_{\rho}\omega_{\rho}+V_{\rho}\omega_{\mu})B_{4} \nonumber \\
&+&\frac{1}{2m^{2}\omega \cd p}
(V_{\mu}p_{\rho}+V_{\rho}p_{\mu}) B_{5}\ ,
\end{eqnarray}
where $V_{\mu}=\varepsilon_{\mu \alpha \beta \gamma} \omega^{\alpha}q^{\beta}p^{\gamma}$. From
  Eq.~(\ref{eq47}), we can extract the physical form factor $F_{\pi \gamma}$ by the
 following contraction
\begin{equation}\label{eq48}
F_{\pi \gamma}=\frac{i}{2Q^{2}(\omega \cd p)} \varepsilon^{\mu \rho \nu \lambda} q_{\nu}
 \omega_{\lambda} F_{\mu \rho}\ .
\end{equation}
For the transition form factor in the impulse approximation, the first relevant diagram, $F_{\mu \rho} ^{a}$,  is indicated in Fig.~\ref{transition}.
By applying  the diagrammatic rules of CLFD, we can derive the corresponding amplitude and get
\begin{eqnarray}\label{eq49}
F_{\mu \rho}^a &=&  \sqrt{\frac{3}{2}} (e_{u}^{2}-e_{d}^{2}) 
\int  \frac{{\rm d}^3{\bf k_1}}{2\varepsilon_{k_1}}\frac{{\rm d} \tau}{\tau -i \epsilon}  \delta (k_2^{ 2}-m^2)\Theta(\omega \cd k_2)\nonumber \\
&\times &\frac{{\rm d} \tau'}{\tau' -i \epsilon}  \delta (k_2^{ \prime2}-m^2)\Theta(\omega \cd k'_2) \\
 &\times & {\rm Tr} \Bigl[-  \overline{\gamma_{\mu}}(\sla k_{2}^{\prime}-
\sla\omega \tau^{\prime} +m) \gamma_{\rho}
(\sla k_{2}+m) {\Gamma_2}(m- \sla k_{1})\Bigr] \ . \nonumber
\end{eqnarray}
The second contribution involving the coupling of the virtual photon to the antiquark can be calculated similarly. Other diagrams which should be taken into account at leading order  either correspond to
 vacuum  diagrams or are equal to zero for $\omega \cd q = 0$. After calculation of the trace, the total amplitude for
 the transition form factor reads
\begin{eqnarray}\label{eq51}
F_{\pi \gamma}(Q^2)&=& 
\frac{4\sqrt{3} (e_{u}^{2}-e_{d}^{2}) }{(2\pi)^3} \int \frac{{\rm d}^{2} {\bf{R}}_{\perp} {\rm d} x}{2x(1-x)}  \nonumber \\
&\times& \frac{x}{m^2 + {\bf {R}}_{\perp}^2 -2 {\bf {R}_{\perp}}
 \cd {\bf \Delta}+x^2Q^2} \\
&\times&\left[A_{1}+2x(1-x)A_{2}- \frac{{\bf {R}_{\perp}} \cd {\bf \Delta}}
{Q^{2}}(1-x) A_{2} \right]\ . \nonumber 
\end{eqnarray}
The transition form factor of the pion is completely finite thanks to the extra dependence on the transverse momentum as compared to the decay constant (\ref{eq38}). The amplitude (\ref{eq49}) includes a contact interaction associated to the elementary quark propagator between the virtual and real photons. It gives the factor $-\, \omega\!\!\!\!/ \tau'$ in this equation. Additional contributions from contact interactions are discussed below.

It is instructive to compare our result (\ref{eq51}), with the one obtained in the asymtotic limit using the pion distribution amplitude \cite{BL}. This can be done  by neglecting the mass term $m^2$ and the transverse momentum squared ${\bf R_\perp}^2$ in  (\ref{eq51}). We recover in this case the standard expression for the transition form factor and its $1/Q^2$ behavior. In our full calculation however, there is no need to regularize our expression in the $Q^2 \to 0$ limit in order to get the low momentum regime \cite{noguera}. 

Comparing this result with the expression of the pion decay constant in (\ref{eq38}), one may naively identify an "equivalent" distribution amplitude given by
\begin{equation} \label{DAeq}
\phi_\pi^{eq}(x)=\frac{1}{F} \frac{2 \sqrt{6}}{(2\pi)^3)}\int \frac{d^2{\bf R_\perp}}{2x(1-x)} \left[ A_1+2x(1-x)A_2\right]\ .
\end{equation}
This "equivalent" distribution amplitude should be compared with the standard asymptotic one $\phi_\pi^{as}=6x(1-x)$ normalized according to 
\begin{equation}
\int \phi_\pi^{as}(x) dx = 1\ .
\end{equation}
This however can not be done safely since the limits $Q^2 \to \infty$ and ${\bf R_\perp}^2 \to \infty$ do not commute for the calculation of the transition form factor in (\ref{eq51}). Indeed, one has to keep the full dependence of the transition form factor as a function of the transverse momentum in order to get a converged result in the limit ${\bf R_\perp}^2 \to \infty$. If we do the limit $Q^2 \to \infty$ by keeping ${\bf R_\perp}^2$ finite, the transition form factor is divergent, similarly to the calculation of the decay constant. This implies also that the "equivalent" distribution amplitude defined in (\ref{DAeq}) is divergent when $\delta A_1$ and $\delta A_2$ are calculated from a one gluon exchange process, as shown in Sec.~\ref{radia}.

\subsection{Contact interactions}
The contribution from one gluon exchange to the physical observables generates in LFD several terms involving contact interactions \cite{brodsky,cdkm}. These ones originate from the singular nature of the LF Hamiltonian. According to the diagrammatic rules given in \cite{cdkm}, one should add, to each fermion (anti-fermion) propagator between two elementary vertices, a contribution of the form $- \omega\!\!\!\!//2\omega\cd k$ ($\sla\omega/2\omega\cd k$), where $k$ is the momentum of the fermion. These contact interactions have been identified in \cite{kdm} to part of the usual meson-exchange currents in the non-relativistic framework. 

\begin{figure}[b]
\begin{center}
\includegraphics[width=8cm]{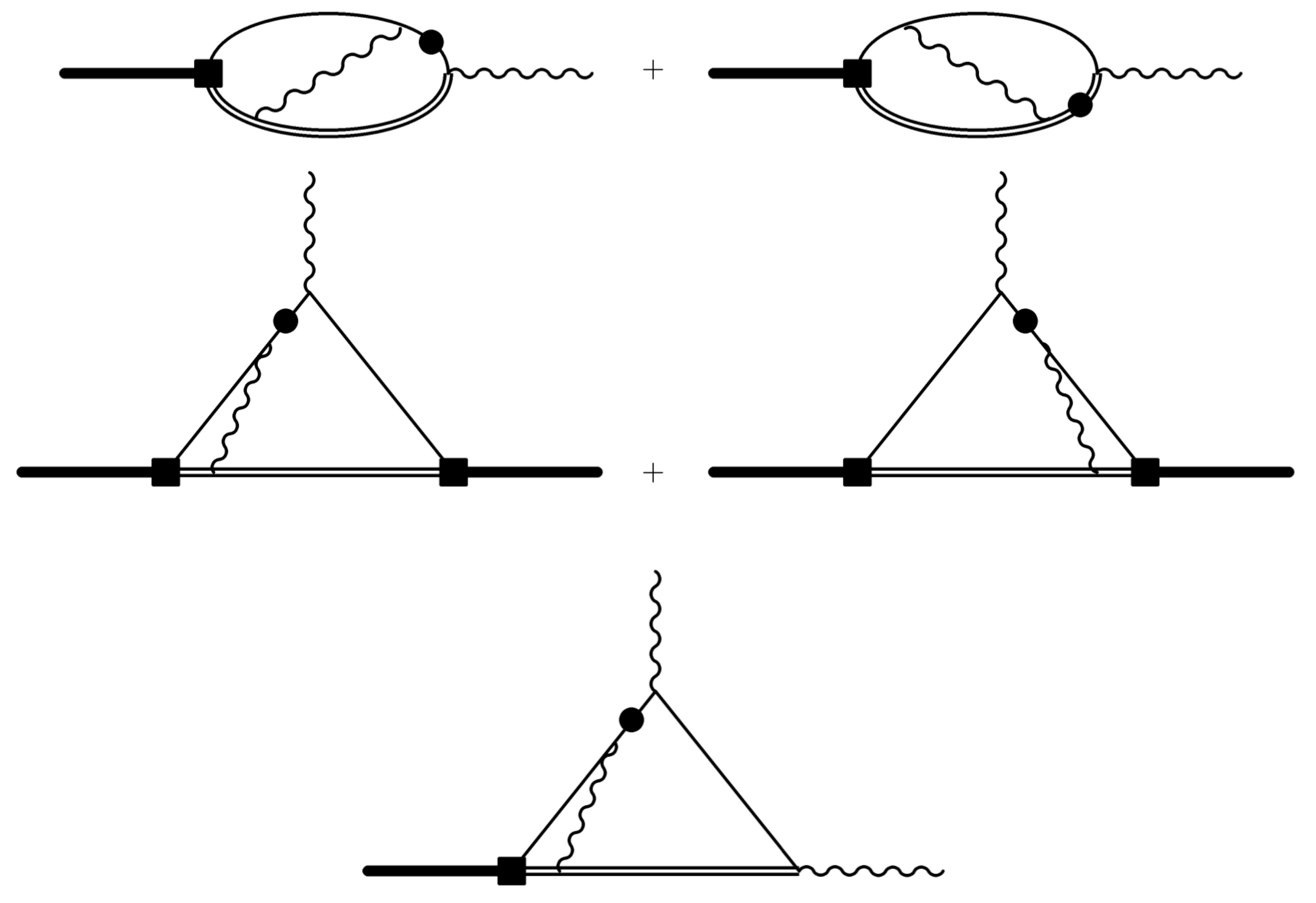}
\end{center}
 \caption{Contribution from the contact interaction (full dot) to the pion decay constant, electromagnetic and transition form factors, from top to bottom respectively, in the impulse approximation.}
 \label{contact}
\end{figure}

For the processes under consideration in this study, we have thus to consider extra contributions to the pion decay constant, electromagnetic and transition form factors. These are shown schematically  in Figs.~\ref{contact}. The contact interaction is indicated by a dot on these figures.
One can easily see that since the contact interaction is proportional to $\omega\!\!\!\!/$, it does not contribute to the pion decay constant and electromagnetic form factor, according to Eqs.~(\ref{eq35},\ref{eq40}). Its contribution to the transition form factor is however very small, at most 1.5\% for the highest measured momentum transfer. It is not included in the numerical results.

%

\section{Numerical results and discussion} \label{numerics}
Our phenomenological analysis has three independent parameters. The first one, $\beta$, gives the typical size of the non-relativistic wave function we start from in Eqs.~(\ref{eq15},\ref{eq15b}). The second parameter is the quark (or antiquark) constituent mass $m$. The third one is the strong coupling constant in the low momentum region given by  $\alpha_s^0$ in Eq.~(\ref{alphas}). The values of these parameters are indicated in Table~\ref{tab1}, for the two types of non-relativistic wave functions used in this study.
\begin{table}[tbhp]
\begin{center}
\begin{tabular}{c|ccc}  
\hline
\hline
                  				& $\beta$    	&   $m$  	 			&   $\alpha_s^0$ 	 \\
\hline
 Gaussienne  w.f.   		&  $3.5$    	&  $250 \mbox{ MeV}$   	&  $1.3$   		\\
Power-law   w.f.    		& $3.72 $      	&  $250 \mbox{ MeV}$	& $0.35$   		\\
\hline
\hline
\end{tabular}
\end{center}
\caption{Parameter sets of the calculation.}
\label{tab1}
\end{table}
\begin{figure*}[bt]
\begin{center}
\includegraphics[width=19.5pc]{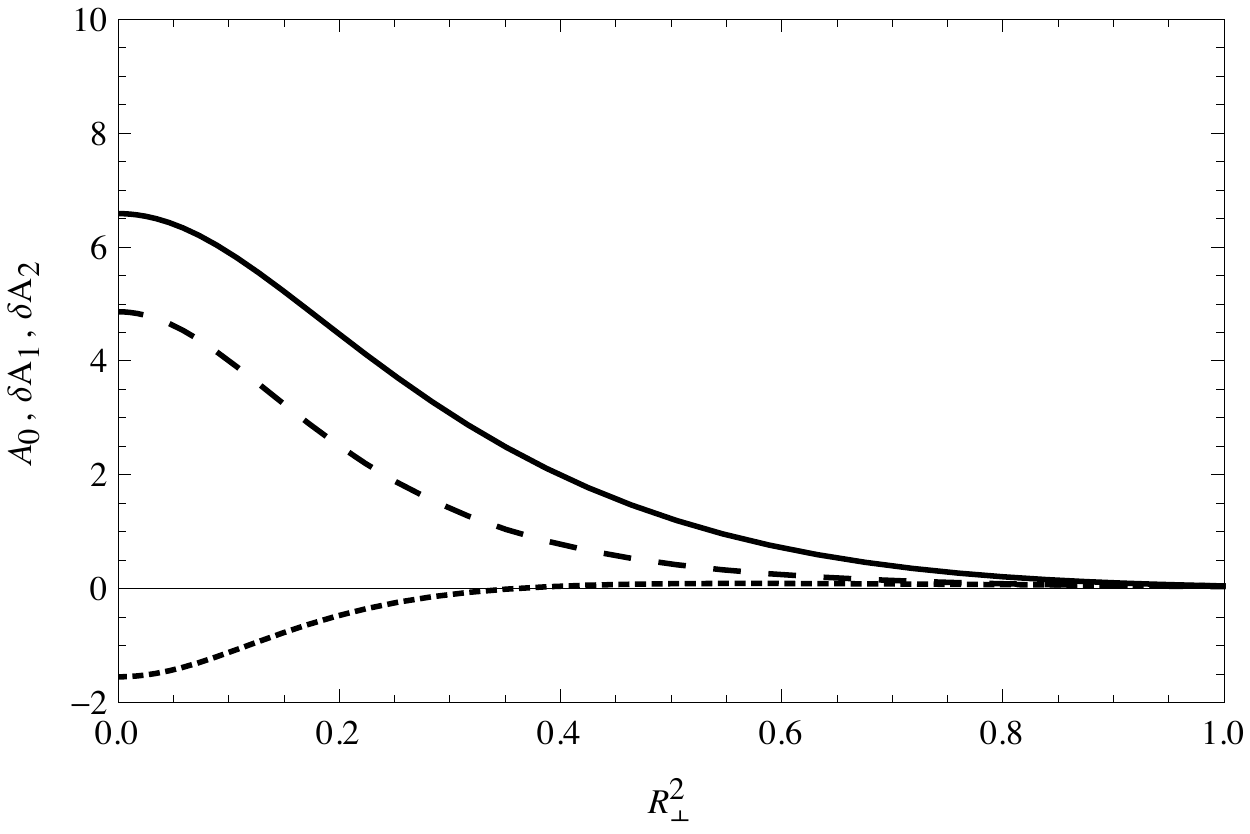}
\includegraphics[width=20pc]{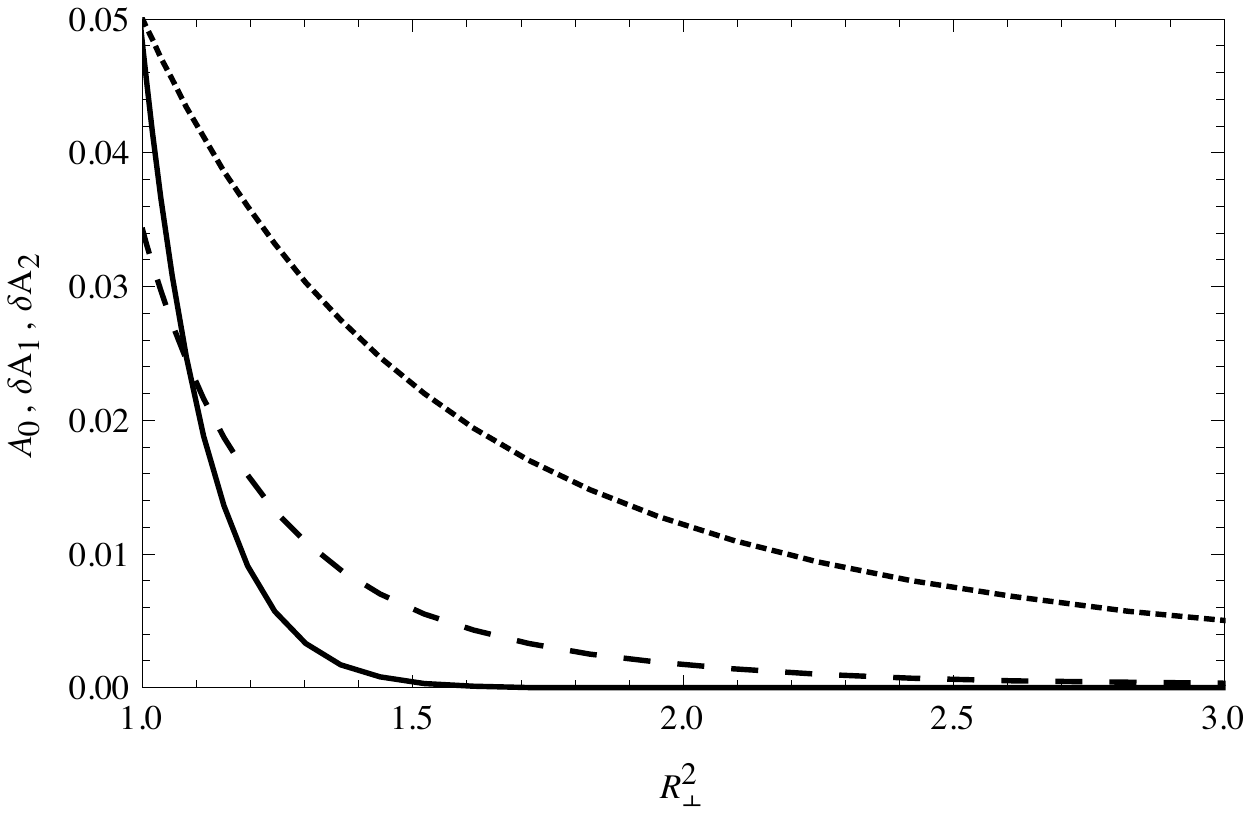}
\end{center}
 \caption{The two components of the pion wave function calculated with a gaussian parametrization in the non-relativistic limit, both in the low (top curve) and high (bottom curve) momentum range. The solid line represents $A_1^0$ in (\ref{O0}) while the dashed (dotted) line represents $\delta A_1$ ($\delta A_2$) in (\ref{do}). The calculation is done for $x=0.5$.}
 \label{wave_gauss}
\end{figure*}
\begin{figure*}[bt]
\begin{center}
\includegraphics[width=19.5pc]{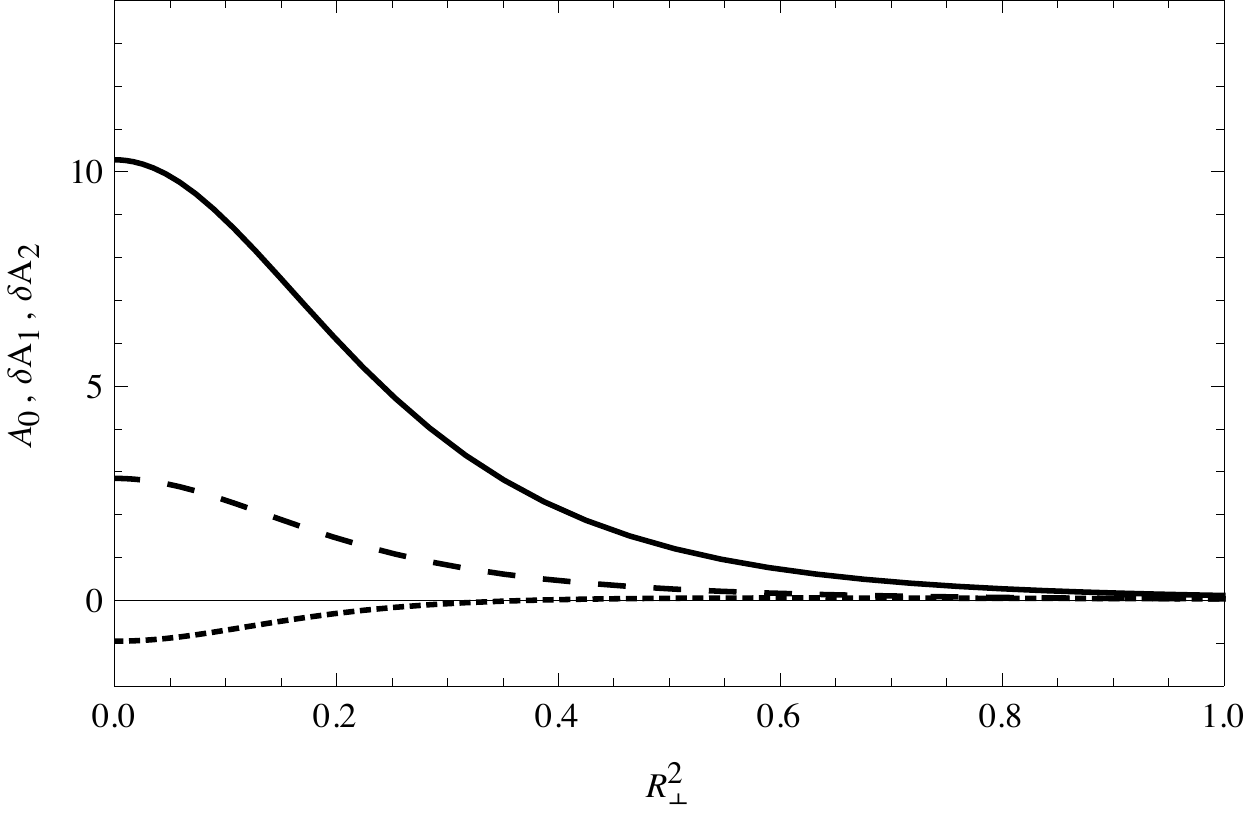}
\includegraphics[width=20pc]{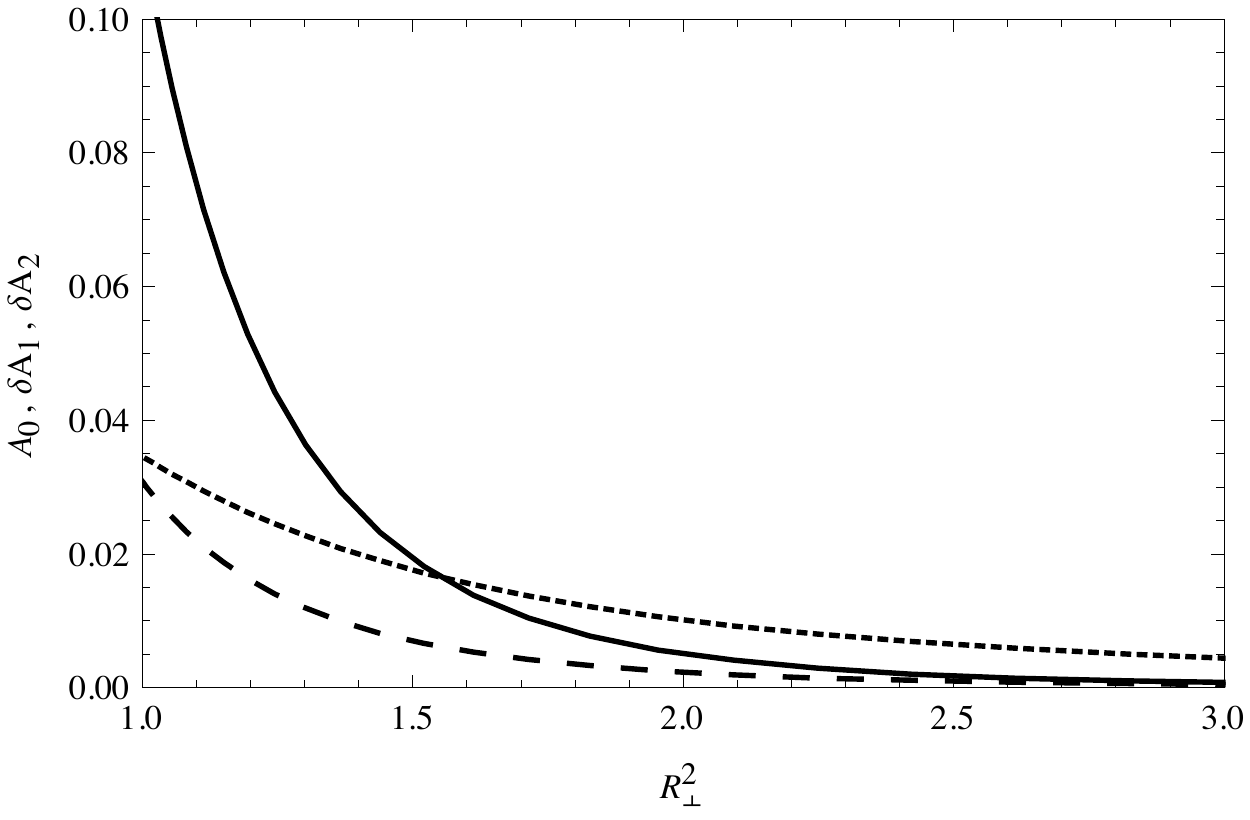}
\end{center}
 \caption{Same as Fig.~(\ref{wave_gauss}) but for the power-law parametrization.}
 \label{wave_pl}
\end{figure*}

These three parameters are fixed to get an overall good description of the pion decay constant, charge radius, electromagnetic and transition form factors. Since the pion decay constant and charge radius are known with a rather good accuracy, we fix two of the parameters to reproduce these quantities (within experimental errors), while the third one is fixed to get an overall good account of the pion electromagnetic and transition form factors at moderate $Q^2$.

We show in Figs.~\ref{wave_gauss}-\ref{wave_pl} the components $A_1$ and $A_2$ of the pion wave function for the two non-relativistic parame-trizations used in this study, in both the low and high momentum range. In the low momentum range, the purely phenomenological component $A_1^0$ dominates. However, the contribution from one gluon exchange given by $\delta A_1$, is of the same order of magnitude for the gaussian parametrization, while it is a factor $2-3$ smaller for the power-law parametrization. This reflects directly the difference in the value of the coupling constant $\alpha_s^0$. The relativistic component $\delta A_2$ is always smaller, but still sizeable.

In the high momentum domain, for ${\bf R_\perp}^2 > 1-2$ GeV$^2$, the relativistic component $\delta A_2$ dominates, as expected from its analytic behavior found in Sec.~\ref{radia}. We clearly see on these figures the interest to take into account the full structure of the pion wave function. It enables to describe, in a unique framework, both the low and high momentum range.
\begin{table}[b,t,h,p]
\begin{center}
\begin{tabular}{c|ccc}  
\hline
\hline
                  	$F$			& Full   	&   $\delta {\cal O} =0$ 	 &  $\delta A_2=0$ 	 \\
\hline
Gaussian  w.f.   		&  $131 $    	& $92 $    		& $140 $    		\\
Power-law   w.f.    		&  $131 $		& $118 $    	&  $149 $     		\\
\hline
\hline
\end{tabular}
\end{center}
\caption{Pion decay constant. All entries are in MeV. }
\label{tab2}
\end{table}
\begin{table}[tbhp]
\begin{center}
\begin{tabular}{c|ccc}  
\hline
           $\langle r_\pi^2 \rangle^{1/2}$       & Full     	&    $\delta {\cal O} =0$	&  $ \delta A_2=0$	 \\
\hline
Gaussian  w.f.   		&  $0.67$    	&  $0.44$   	&  $0.68$    		\\
Power-law   w.f.    		&  $0.67$    	&  $0.54$   	&  $0.68$    		\\
\hline
\hline
\end{tabular}
\end{center}
\caption{Pion charge radius. All entries are in fm.}
\label{tab3}
\end{table}

\begin{figure*}[bt]
\begin{center}
\includegraphics[width=8.5cm]{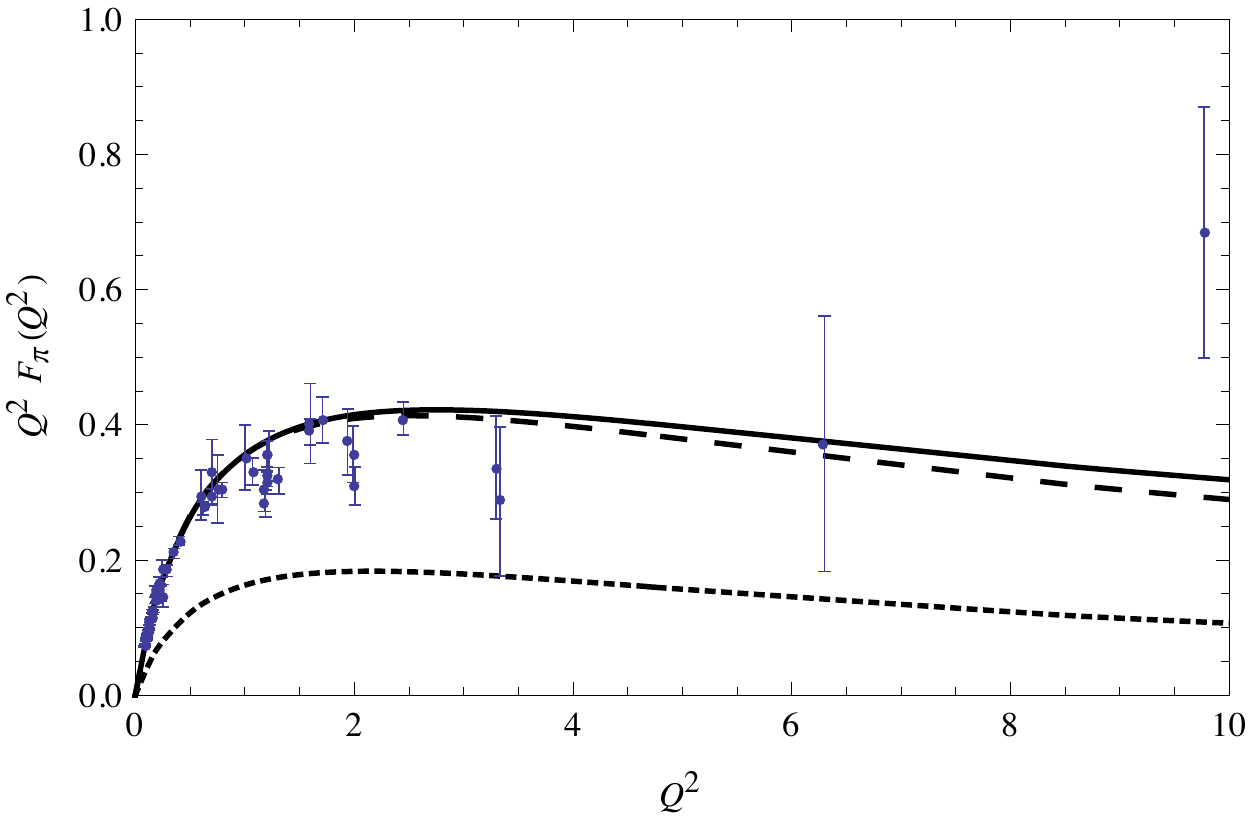}
\includegraphics[width=8.5cm]{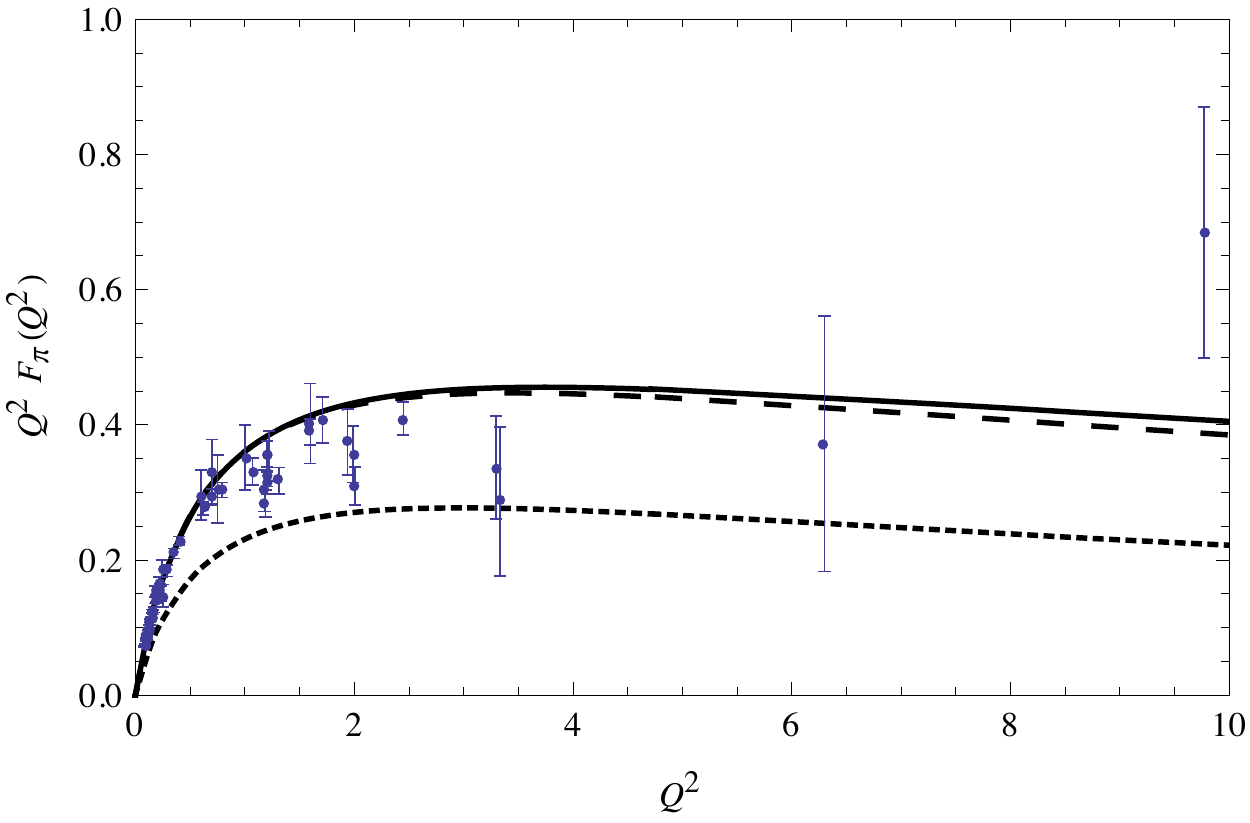}
\end{center}
 \caption{Pion electromagnetic form factor calculated with a gaussian (left plot) and a power-law (right plot) wave functions in the non-relativistic limit. The solid line is the complete calculation, the dotted line is the calculation without any correction from one gluon exchange (i.e. with $\delta A_1=\delta A_2=0$), while the dashed line corresponds to  $\delta A_2=0$. The experimental data  are from \cite{pion1,pion2,pion3,pion4,pion5,pion6,pion7,pion8}.}
 \label{ffm}
\end{figure*}

\begin{figure*}[bt]
\begin{center}
\includegraphics[width=8.5cm]{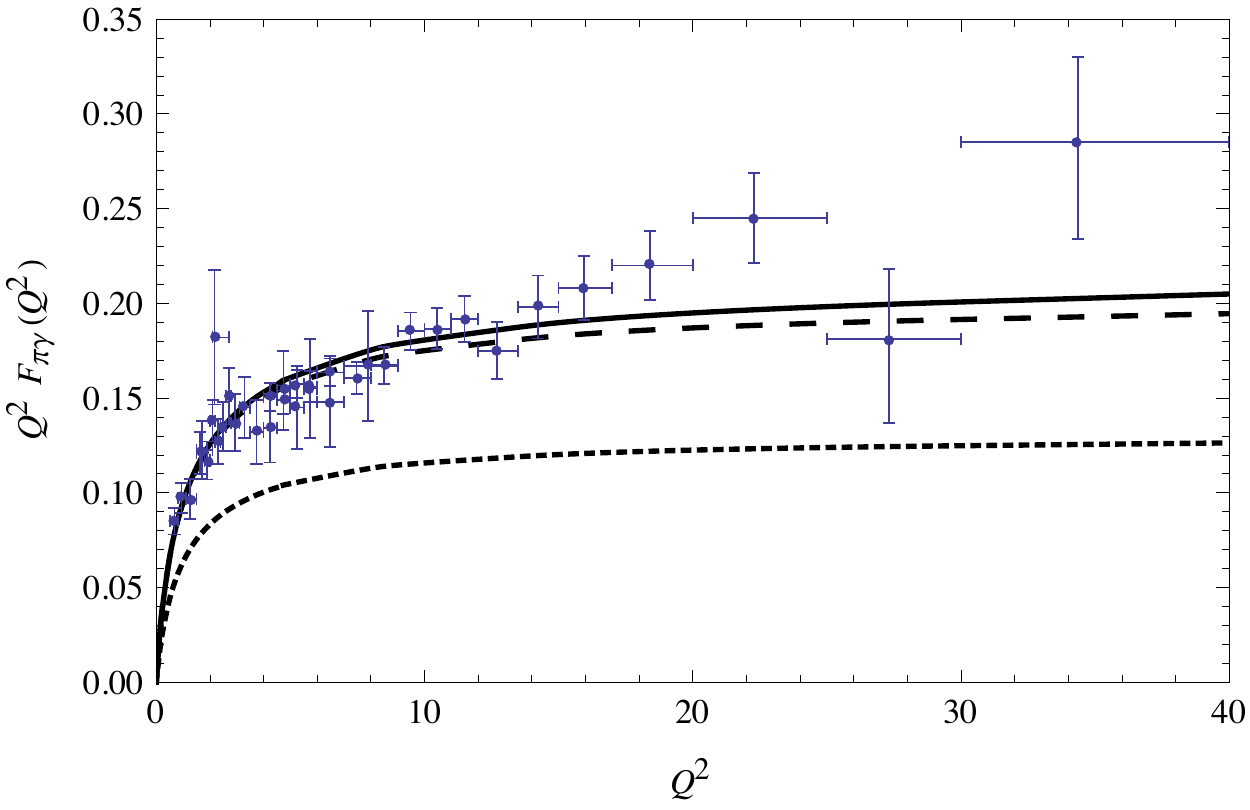}
\includegraphics[width=8.5cm]{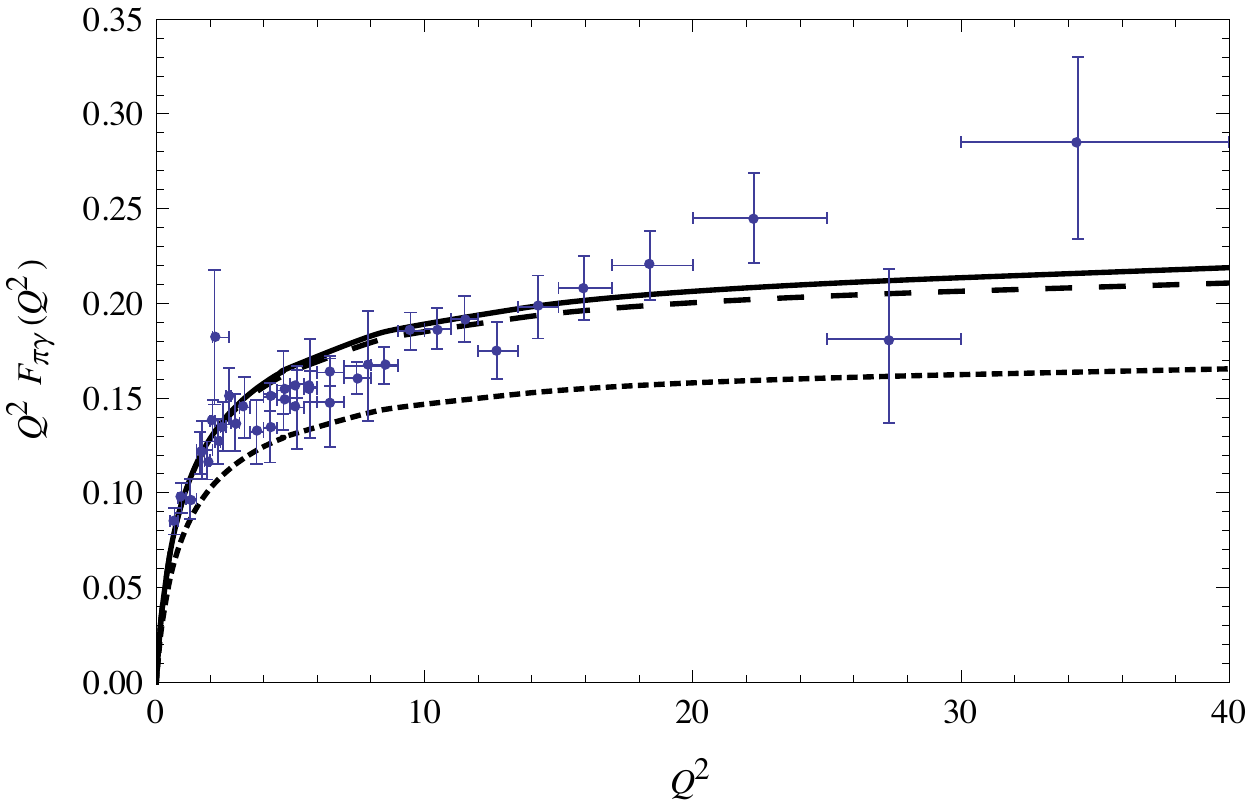}
\end{center}
 \caption{Pion transition form factor calculated with a gaussian (left plot) and a power-law (right plot) wave functions in the non-relativistic limit. The solid line is the complete calculation, the dotted line is the calculation without any correction from one gluon exchange (i.e. with $\delta A_1=\delta A_2=0$), while the dashed line corresponds to  $\delta A_2=0$. The experimental data are from \cite{transition3,transition1,transition2}.}
 \label{fft}
\end{figure*}

Our predictions for the pion decay constant and charge radius are shown in Tables~\ref{tab2} and \ref{tab3}, respectively. The electromagnetic and transition from factors are shown in Figs.~\ref{ffm}-\ref{fft} for the two types of non-relativistic wave functions used in this study. Given the large experimental errors at large momentum transfer, we do not attempt in this study to get a best fit to all the data, but just to show that an overall agreement of all the available data is possible within our framework.

The pion electromagnetic form factor is shown in  Figs. \ref{ffm} together with the world-wide experimental data. Given the experimental errors which are large above $3$ GeV$^2$, both parametrization (gaussian or power-law) give a rather good account of the data, in the whole kinematical domain available. 

In order to settle the importance of the various components of the pion wave function, we also show in these figures the electromagnetic form factor calculated with $\delta A_2=0$ (dashed line). In the kinematical domain $Q^2< 10$ GeV$^2$, the contribution of the component $A_2$ is rather small. This may be surprising given that $A_2$ dominates the wave function for ${\bf R_\perp}^2 > 2$ GeV$^2$, as shown in Figs.~\ref{wave_gauss}, \ref{wave_pl}. This indicates that the $Q^2$ domain where the asymptotic regime is dominant, i.e. where $A_2$ dominates in the calculation of the electromagnetic form factor according to (\ref{eq42}), is very high. This is in full agreement with the early discussions in Refs.~\cite{isgur}. With our numerical parameters, it is above $100$ GeV$^2$, much above the present experimental data. At moderate $Q^2$, both low and moderate momentum domain of $A_2$ dominate, and there is a partial cancellation between these contributions from the change in sign of $\delta A_2$ at about $0.5$ GeV$^2$.

The dotted line on these figures shows the contribution of $A_1^0$ only. It is sizeably smaller than the full calculation. This originates directly from the importance of the $\delta A_1$ contribution to  $A_1$ component, as shown on Figs.~\ref{wave_gauss}-\ref{wave_pl}. Note that the complete calculations using the gaussian or power-law parametrizations are extremely similar, the only difference being in the value of $\delta A_1$ and $\delta A_2$. This may indicate that the most important feature, in this kinematical domain, is to have enough high momentum components in the pion wave function, either in the $A_1$ or in the$A_2$ components. It can come from the non-relativistic parametrization $A_1^0$ or from the one gluon exchange process giving rise to $\delta A_1$ and $\delta A_2$. Since the gaussian parametrization has very little high momentum components, this should be compensated by a larger $\alpha_s^0$.

The corresponding results for the pion transition form factor are shown on Figs.~\ref{fft}. The qualitative, and to some extend also quantitative, features we get in this case are very similar to the ones detailed above for the electromagnetic form factor. At very high momentum transfer however, for $Q^2> 15$ GeV$^2$, our results underestimate slightly the experimental data, with a better agreement when using a power-law wave function. There is no way to adjust our parameters to get a better agreement for the transition form factor without spoiling the good agreement we get for the electromagnteic form factor. We should however wait for more precise experimental data before drawing any definite conclusions.

\section{Summary and discussion} \label{conc}
We have investigated in this study the full relativistic structure of the pion in the framework of the constituent quark model. This structure involves two spin components, which, in turn, depends on two kinematical variables, like for instance the longitudinal momentum fraction and the square of the transverse momentum. This complete calculation has been made possible by the use of the explicitly covariant formulation of light-front dynamics \cite{cdkm}. Our phenomenological analysis has been compared with the full set of observables available at present: the pion decay constant, the charge radius, the electromagnetic and the transition form factors. These observables involve both low and high momentum scales.

Our wave function is constructed starting from a purely phenomenological wave function in the non relativistic limit. Relativistic kinematical corrections are thus included exactly using CLFD, while dynamical relativistic corrections are included by a one gluon exchange process. The latter generates the necessary relativistic high momentum components in the pion wave function.

From this full structure of the pion wave function, we have been able to obtain an overall very good agreement with all experimental data available, both in the low and high momentum domain. To get more physical insight into the relevant components of the wave function, it is however necessary to have more precise measurements of the pion electromagnetic form factor in the momentum range above $Q^2\simeq 5 \mbox{ GeV}^2$. It is also necessary to confirm the recent Babar data for the pion transition form factor at very high momentum transfer (till about $Q^2\simeq 40 \mbox{ GeV}^2$), with more precise data.

This analysis shows also the real flexibility of CLFD in describing few body systems in relativistic nuclear and particle physics. Its application to more fundamental calculations starting from first principles is also under way \cite{LC2010}.

%
\subsubsection*{Acknowledgements}
%
One of us (O.L.)  would like to thank X.-H. Guo for useful and stimulating
 discussions. We also thank V. Karmanov for fruitful discussions and comments about
 this work.


\begin{thebibliography}{}
\bibitem{transition3}
B. Aubert {\it et al.}, Phys. Rev. {\bf D80} (2009) 052002.
\bibitem{transition1}
H.J. Behrend {\it et al.}, Z. Phys. {\bf C49} (1991) 401.
\bibitem{transition2}
J. Gronberg {\it et al.}, Phys. Rev. {\bf D57} (1998) 33.
\bibitem{pion1}
C.N. Braun {\it et al.}, Phys. Rev. {\bf D8} (1973) 92.
\bibitem{pion2}
C.J. Bebek {\it et al.},Phys. Rev. {\bf D17} (1978) 1693.
\bibitem{pion3}
H. Ackermann {\it et al.}, Nucl. Phys. {\bf B137} (1978) 294.
\bibitem{pion4}
P. Brauel {\it et al.}, Z. Phys. {\bf C3} (1979) 101.
\bibitem{pion5}
S.R. Amendolia {\it et al.}, Phys. Lett. {\bf B178} (1986) 435; Phys. Lett. {\it B277} (1986) 168.
\bibitem{pion6}
J. Volmer {\it et al.}, Phys. Rev. Lett. {\bf 86} (2001) 1713.
\bibitem{pion7}
T. Horn {\it et al.}, Phys. Rev. Lett. {\bf 97} (2006) 192001.
\bibitem{pion8}
V. Tadevosyan {\it et al.}, Phys. Rev. {\bf C75} (2007) 055205.
\bibitem{pdg}
Review of Particle Physics, Phys. Lett. {\bf B667} (2008) 1.
\bibitem{BL}
G.P. Lepage and S.J. Brodsky, Phys. Rev. {\bf D22} (1980) 2157.
\bibitem {mikhailov}
S. V. Mikhailov, N.G. Stefanis, Nucl. Phys. {\bf B821} (2009) 291
\bibitem{noguera}
S. Noguera and V. Vento, {\it '' The pion transition form factor and the pion distribution amplitude''} arXiv: 1001.3075
\bibitem{bakulev}
A.P. Bakulev, S.V. Mikhailov, and N.G. Stefanis, Phys. Lett. {\bf B508} (2001) 279.
\bibitem{li}
H.N. Li and G. Sterman, Nucl. Phys. {\bf B381} (1992) 129.
\bibitem{yakovlev}
A. Schmedding and O.I. Yakovlev, Phys. Rev. {\bf D62} (2000) 116002.
\bibitem{cao}
F.G. Cao, T. Huang and B.Q. Ma, Phys. Rev. {\bf D 53} (1996) 6582.
\bibitem{musatov}
I.V. Musatov and A.V. Radyushkin, Phys. Rev. {\bf D56} (1997) 2713.
\bibitem{raha}
U. Raha and A. Aste, Phys. Rev. {\bf D79} (2009) 034015.
\bibitem{jakob}
R. Jakob and P. Kroll, Phys. Lett. {\bf B 315} (1993) 463.
\bibitem{jkr}
R. Jakob, P. Kroll, and M. Raulfs, J. Phys. {\bf G22} (1996) 45.
\bibitem{hww}
T. Huang, X.G. Wu and X.H. Wu, Phys. Rev. {\bf D70} (2004) 053007.
\bibitem{Dirac}
P.~A.~M.~Dirac, Rev. Mod. Phys. {\bf 21}, 392 (1949).
\bibitem{karm76}
V.A.~Karmanov, Zh. Eksp. Teor. Fiz.  {\bf 71}, 399 (1976);
[transl.: Sov. Phys. JETP 44, 210 (1976)].
\bibitem{brodsky}
S. J. Brodsky, H.-C. Pauli, S. Pinsky, Phys. Reports {\bf 301} (1998) 299.
\bibitem{cdkm}
J.~Carbonell, B.~Desplanques, V.A.~Karmanov and J.-F.~Mathiot,
Phys. Reports   {\bf 300}, 215 (1998).
\bibitem{dugne}
F. Bissey, J.-J. Dugne, J.-F. Mathiot, Eur. Phys. J. {\bf C24} (2002) 101
\bibitem{kroll}
P. Kroll and H. Raulfs, Phys. Lett. {\bf B 387} (1996) 848.
\bibitem{belyaev}
V.M. Belyaev and M.B. Johnson, Phys. Rev. {\bf D56} (1997) 1481.
\bibitem{schlumpf}
F. Schlumpf, Phys. Rev. {\bf D50} (1994) 6895.
\bibitem{BHL}
S.J. Brodsky, T. Huang and P. Lepage, in {\it ''Particles and Fields''}, A.Z. Capri and A.N. Kamal, Eds, Plenum Publishing Corporation, New-York 1983.
 \bibitem{kms_08}
V.A.~Karmanov, J.-F.~Mathiot and A.V.~Smirnov, Phys. Rev. {\bf
D77}, 085028 (2008).
\bibitem{kdm}
V.A. Karmanov, B. Desplanques and J.-F. Mathiot, Nucl. Phys. {\bf A589} (1995) 697
\bibitem{isgur}
N. Isgur and C.H. Llewellyn Smith, Phys. Rev. Lett. {\bf 52} (1984) 1080; Phys. Lett. {\bf B217} (1989) 535; Nucl. Phys. {\bf B317} (1989) 526.

\bibitem{LC2010}
J.-F. Mathiot, {\it Field theory on the light front}, contribution to the conference "LC2010: relativistic hadronic and particle physics", Valencia (Spain), June 2010, to be published in Proceedings of Science.
\end{thebibliography}
\end{document}